\documentclass[prb,twocolumn,superscriptaddress,showpacs]{revtex4}
\usepackage{amsmath,amssymb,graphicx,color,bm,epstopdf}

\begin{document}
\title{Band gap in graphene induced by vacuum fluctuations}

\author{O. V. Kibis}\email{Oleg.Kibis@nstu.ru}

\affiliation{Department of Applied and Theoretical Physics,
Novosibirsk State Technical University, Karl Marx Avenue 20,
630092 Novosibirsk, Russia} \affiliation{International Institute
of Physics, Av. Odilon Gomes de Lima, 1772, Capim Macio,
59078-400, Natal, Brazil}

\author{O. Kyriienko}
\affiliation{Science Institute, University of Iceland, Dunhagi-3,
IS-107, Reykjavik, Iceland}

\author{I. A. Shelykh}
\affiliation{Science Institute, University of Iceland, Dunhagi-3,
IS-107, Reykjavik, Iceland} \affiliation{International Institute
of Physics, Av. Odilon Gomes de Lima, 1772, Capim Macio,
59078-400, Natal, Brazil}


\begin{abstract}
The electrons in undoped graphene behave as massless Dirac
fermions. Therefore graphene can serve as an unique
condensed-matter laboratory for the study of various relativistic
effects, including quantum electrodynamics (QED) phenomena.
Although theoretical models describing electronic properties of
graphene have been elaborated in details, the QED effects were
usually neglected. In this paper we demonstrate theoretically that
QED can drastically modify electronic properties of graphene. We
predict the following QED effect
--- the opening of the band gap in a graphene monolayer placed
inside a planar microcavity filled with an optically active media.
We show that this phenomenon occurs due to the vacuum fluctuations
of the electromagnetic field and is similar to such a well-known
phenomenon as a vacuum-induced splitting of atomic levels (the
Lamb shift). We estimate the characteristic value of the band gap
and find that it can sufficiently exceed the value of the Lamb
shift.
\end{abstract}


\maketitle


\section{Introduction} Graphene --- a monolayer of carbon atoms ---
possesses unusual physical properties that make it attractive for
various applications.\cite{Geim,Geim_07,Kostya}  Usually treated
as a platform for the novel high-speed
electronics,\cite{Morozov,Sarma_11} graphene is of great interest
from the point of view of the fundamental physics as well. Indeed,
the low-energy electron excitations in graphene are massless Dirac
fermions with the linear energy spectrum,
$\varepsilon(\mathbf{k})=\hbar
v_F|\textbf{k}|$.\cite{NovoselovNature,CastroNetoRMP,Lukyanchuk,Semenoff}
That makes graphene a condensed-matter playground for the study of
various relativistic quantum phenomena, such as the Klein
tunnelling \cite{Beenakker,Katsnelson} and the Casimir
effect.\cite{Dutra,Fialkovsky_11,Sernelius_11} Up to now, most of
graphene-related studies were focused on its unusual transport
properties, and quantum electrodynamics (QED) effects arising from
interaction of electrons in graphene with a quantized
electromagnetic field were neglected.\cite{CastroNetoRMP} This
paper is aimed to fill partially this gap in the theory. We show
that due to the giant Fermi velocity of electrons in graphene,
$v_{F}\approx c/300$, QED effects are pronounced and can lead to
qualitative modifications of the spectrum of elementary
excitations.

The linear energy spectrum of electrons in graphene comes from its
specific honeycomb lattice structure which makes the band gap
between the valence and conductivity bands to be exactly
zero.\cite{CastroNetoRMP} There is the long-standing problem of
the opening of the band gap. The appearance of a controllable band
gap is required for various electronical and optical applications
of graphene.\cite{Geim,Avouris,Hartmann} Aside from this, it is
interesting from the fundamental viewpoint to analyze how massless
Dirac fermions can acquire a mass. This question is relevant,
particularly, in the context of the observation of Majorana
fermions in condensed matter systems.\cite{Wilczek} Several
mechanisms of the band gap opening in monolayer graphene have been
proposed. Among them are breaking of the symmetry between two
sublattices of the honeycomb lattice of
graphene,\cite{Zhou,Guinea} the spin-orbit coupling \cite{Kane}
and the many-body interactions leading to the excitonic
instability.\cite{Gamayun,Gamayun_PRB}

Recently, one of us put forward the proposal of opening the band
gap by illuminating graphene with a circularly polarized
light.\cite{Kibis_10} In this case the gap in the spectrum of
elementary electron excitations appears due to the formation of
composite electron-photon states which are similar to polaritons
in ionic crystals and quantum
microcavities.\cite{KavokinBook,PolaritonDevices,ShelykhReview,Liew}
It should be noted that, within the framework of QED, the
electron-photon interaction can be observed even if ``real''
photons are absent and electrons interact only with vacuum
fluctuations of electromagnetic field due to emitting and
reabsorbing virtual photons.\cite{Landau_4} Therefore, one can
expect that the photon-induced splitting of valence and
conductivity bands in graphene \cite{Kibis_10} will take place due
to the vacuum fluctuations even in the absence of an external
field pumping. This QED effect is similar to the well-known Lamb
shift in the atomic physics, i.e., the vacuum-induced splitting of
the states $2s_{1/2}$ and $2p_{1/2}$ of a hydrogen atom with the
characteristic splitting energy $\Delta\approx 4$ $\mu$eV. The
Lamb shift, discovered experimentally by Lamb and Retherford
\cite{Lamb_47} and theoretically explained by Bethe
\cite{Bethe_47} more than 60 years ago, is extremely important for
understanding and verification of basic principles of QED. That is
why it attracts the undivided attention of the physics community
up to now.\cite{Scully_10}

Since clockwise and counterclockwise polarized photons shift
electron levels in graphene in mutually opposite
directions,\cite{Kibis_10} the band gap opening needs breaking the
symmetry between virtual photons with different circular
polarizations. This can be achieved by placing a graphene
monolayer inside a planar cavity filled with an optically active
material (see Fig. \ref{Fig1}). As it will be shown below, in this
case the vacuum fluctuations lead to the opening of the band gap
in graphene even in the absence of an external circularly
polarized optical pumping. It should be noted, that the QED mass
renormalization in an optically active media has been
considered,\cite{Marzlin} but, surprisingly, the most interesting
case of massless Dirac fermions was not analyzed before.
\begin{figure}
\includegraphics[width=1.0\linewidth]{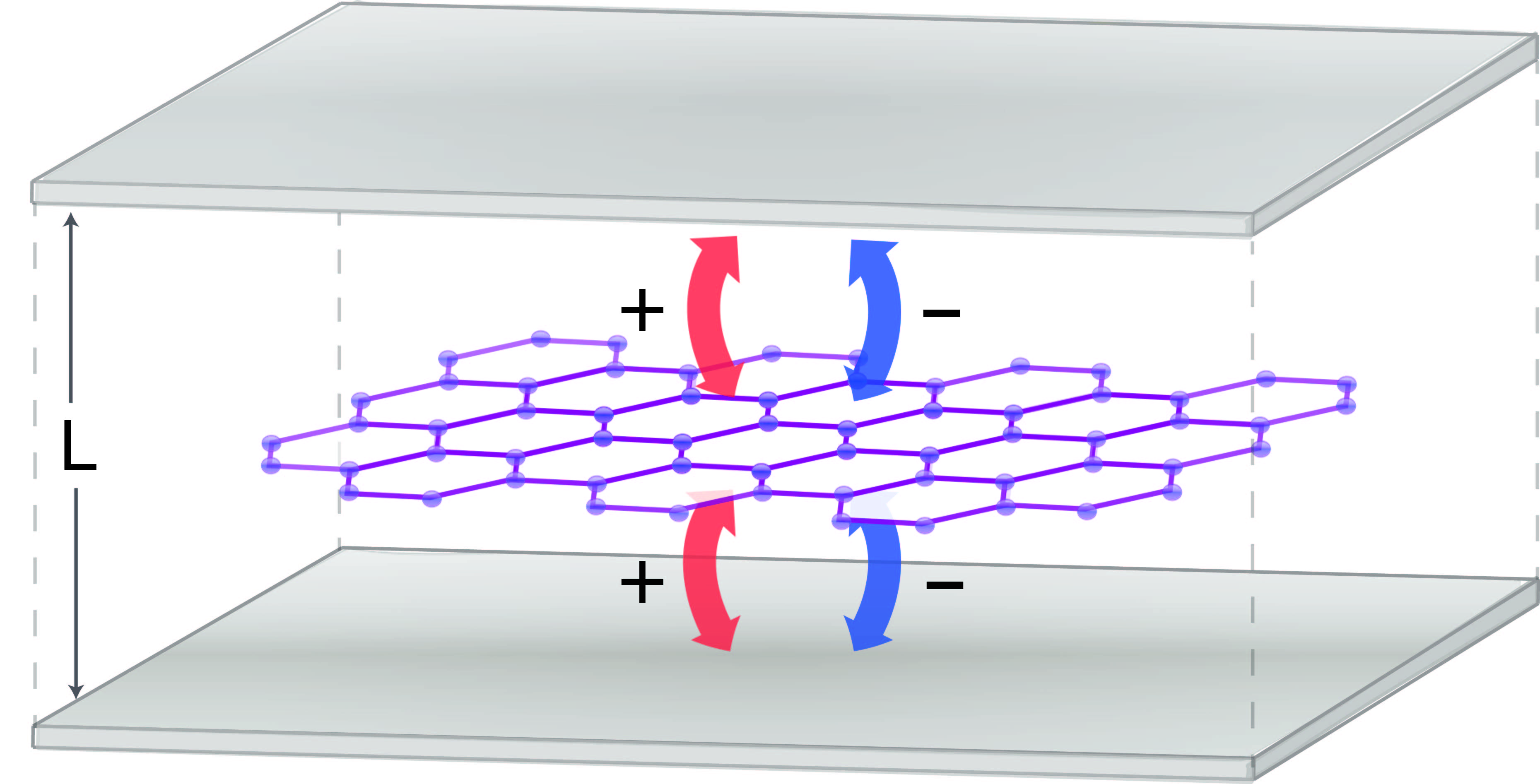}
\caption{(color online) Sketch of the system. A graphene sample
placed inside a planar cavity filled with an optically active
media. The arrows with signs $+$ and $-$ correspond to clockwise
and counterclockwise circularly polarized virtual photons,
respectively, which are emitted and reabsorbed by electrons in
graphene.} \label{Fig1}
\end{figure}

\section{The model} Let us consider the problem of interaction
between a single electron in graphene and a single photon mode of
a planar microcavity. Generally, electron states in graphene near
the Fermi energy are described by the eight-component wave
function which accounts for two elementary sublattices of
graphene, two electron valleys, and two orientations of electron
spin.\cite{CastroNetoRMP} In what follows intervalley scattering
processes and spin-flip effects will be beyond consideration,
which reduces the number of necessary components of wavefunction
to two.

The single-particle Hamiltonian of electron in graphene coupled to
the cavity mode reads (see the Appendix A for details of the
derivation)
\begin{equation}\label{H}
\hat{{\cal H}}=\hat{{\cal H}}_{\mathrm{field}}+\hat{{\cal
H}}_{\mathbf{k}}+\hat{{\cal H}}_{\mathrm{int}}\,,
\end{equation}
where
\begin{equation}\label{HF} \hat{{\cal
H}}_{\mathrm{field}}=\sum_{\mathbf{q},_\pm}\hbar\omega_{\mathbf{q},_\pm}\hat{a}^\dagger_{\mathbf{q},_\pm}\hat{a}_{\mathbf{q},_\pm}
\end{equation}
is the photonic part of the Hamiltonian written in the basis of
circularly polarized states,
\begin{equation}\label{HK}
\hat{{\cal H}}_{{\mathbf{k}}}=\hbar
v_F\hat{\bm{\sigma}}\cdot\mathbf{k}
\end{equation}
is the electron Hamiltonian near the point where the valence and
conductivity bands of graphene touch each other (the Dirac point)
and
\begin{widetext}
\begin{equation}\label{H0}
\hat{{\cal H}}_{\mathrm{int}}=-ev_{F}\sqrt{\frac{2\hbar
}{\epsilon_{0}LS}}\sum_{\mathbf{q}}\left[\frac{1}{\sqrt{\omega_{+,\textbf{q}}}}\left(\hat{\sigma}^{-}\hat{a}_{+,\mathbf{q}}e^{i\textbf{qr}}+
\hat{\sigma}^{+}\hat{a}_{+,\mathbf{q}}^{\dagger}e^{-i\textbf{qr}}\right)+
\frac{1}{\sqrt{\omega_{-,\textbf{q}}}}\left(\hat{\sigma}^{-}\hat{a}_{-,\mathbf{q}}^{\dagger}e^{-i\textbf{qr}}+
\hat{\sigma}^{+}\hat{a}_{-,\mathbf{q}}e^{i\textbf{qr}}\right)\right]
\end{equation}
\end{widetext}
is the Hamiltonian of electron-photon interaction in the cavity.
For definiteness, we assume the graphene sheet to be placed in the
center of the cavity. In Eqs. (\ref{HF})--(\ref{H0}) the subscript
indices, $\pm$, correspond to the photon modes with clockwise and
counterclockwise circular polarizations,
$\mathbf{k}=\textbf{e}_xk_x+\textbf{e}_yk_y$ and
$\mathbf{q}=\textbf{e}_xq_x+\textbf{e}_yq_y$ denote in-plane
electron and photon wave vectors, respectively, $\textbf{e}_{x,y}$
are unit vectors directed along the $x,y$-axis, $e$ is the
electron charge, $\epsilon_0$ is the vacuum permittivity, $L$ is
the distance between two mirrors of the planar cavity (the cavity
length), $S$ is the area of graphene sample,
$\omega_{\pm,\textbf{q}}$ are the eigenfrequencies of clockwise
and counterclockwise circularly polarized photons,
$\hat{a}_{\pm,\textbf{q}}$ and $\hat{a}_{\pm,\textbf{q}}^\dagger$
are photonic annihilation and creation operators. The Pauli vector
operator, $\hat{\bm{\sigma}}$, acts in the space of two orthogonal
electron states, $|\pm\rangle$, corresponding to the two
sublattices of graphene in accordance with the following rules:
$\hat{\sigma}_z|\pm\rangle=\pm|\pm\rangle$ and
$\hat{\sigma}^\pm|\mp\rangle=|\pm\rangle$, where
$\hat{\sigma}^{\pm}=(\hat{\sigma}_x\pm i\hat{\sigma}_y)/2$. Thus,
it corresponds to the \emph{pseudospin} of electron.

Eigenstates of the electron Hamiltonian (\ref{HK}) are given by
the expression \cite{CastroNetoRMP}
\begin{equation}\label{k}
|\mathbf{k},\pm\rangle=\frac{e^{i\mathbf{k}\mathbf{r}}}{\sqrt{2S}}\left(e^{-i\theta_{\mathbf{k}}/2}|+\rangle\pm
e^{i\theta_{\mathbf{k}}/2}|-\rangle\right),
\end{equation}
where $\theta_\textbf{k}=\text{arctan}(k_y/k_x)$ and the signs
$\pm$ correspond to electron states in the conductivity and
valence bands of graphene (the upper and lower Dirac cones,
respectively). The corresponding eigenenergies are
$\varepsilon^{(0)}_{\pm,\textbf{k}}=\pm\hbar v_F|\mathbf{k}|$. The
eigenstates of the photon Hamiltonian (\ref{HF}) can be written as
$|N_{\pm,\textbf{q}}\rangle$, where $N_{\pm,\textbf{q}}$ are
photon occupations number for photons with different circular
polarizations ($\pm$) and wave vectors $\textbf{q}$. Then
eigenstates of the full electron-photon Hamiltonian (\ref{H}) can
be decomposed in the basis of the orthogonal electron-photon
states
\begin{equation}\label{b}
|\mathbf{k},\pm,N_+,N_-\rangle=|\mathbf{k},\pm\rangle\otimes
|N_{+,\textbf{q}}\rangle\otimes|N_{-,\textbf{q}^\prime}\rangle
\end{equation}
with the energies
\begin{equation}\label{e}
\varepsilon^{(0)}_{\textbf{k},\pm,N_+,N_-}=\pm\hbar
v_F|\mathbf{k}|+\hbar\omega_{+,\textbf{q}}N_{+,\textbf{q}}+
\hbar\omega_{-,\textbf{q}^\prime}N_{-,\textbf{q}^\prime}\,.
\end{equation}
In order to find eigenstates and eigenenergies of the full
Hamiltonian (\ref{H}), we will use the perturbation theory,
considering the interaction Hamiltonian (\ref{H0}) as a
perturbation. To calculate the energy corrections in the lowest
order of the perturbation, one needs to find the eigenvalues of
the $2\times2$ matrix $\widetilde{\mathbf{{\cal H}}}^{(1)}$ having
matrix elements $\widetilde{\mathbf{{\cal
H}}}_{ss'}^{(1)}(\textbf{k})=\langle\textbf{k},s,0,0|\hat{{\cal
H}}_{\mathrm{int}}|\textbf{k},s',0,0\rangle$. However, it is easy
to see that all matrix elements of this type are zero and one
needs to use the second order of the perturbation theory.
Physically, we need to account for the following processes: the
electron with a wave vector $\textbf{k}$ emits a virtual photon
with a momentum $\textbf{q}$ and then reabsorbs it. Note, that in
such a process the momentum of the electron in the initial state,
$\textbf{k}$, should be equal to its momentum in the final state,
but the value of the index $s$ can be changed: the electron can
remain in the same Dirac cone or move from one Dirac cone to
another one. The last process becomes efficient around
$\textbf{k}=0$ point where the energies of the Dirac cones are
close to each other, which can lead to the lifting of the
degeneracy as we show below. Therefore, to calculate the spectrum
of the Hamiltonian (\ref{H}), we need to use the perturbation
theory for degenerate states.

\begin{figure}
\includegraphics[width=1.0\linewidth]{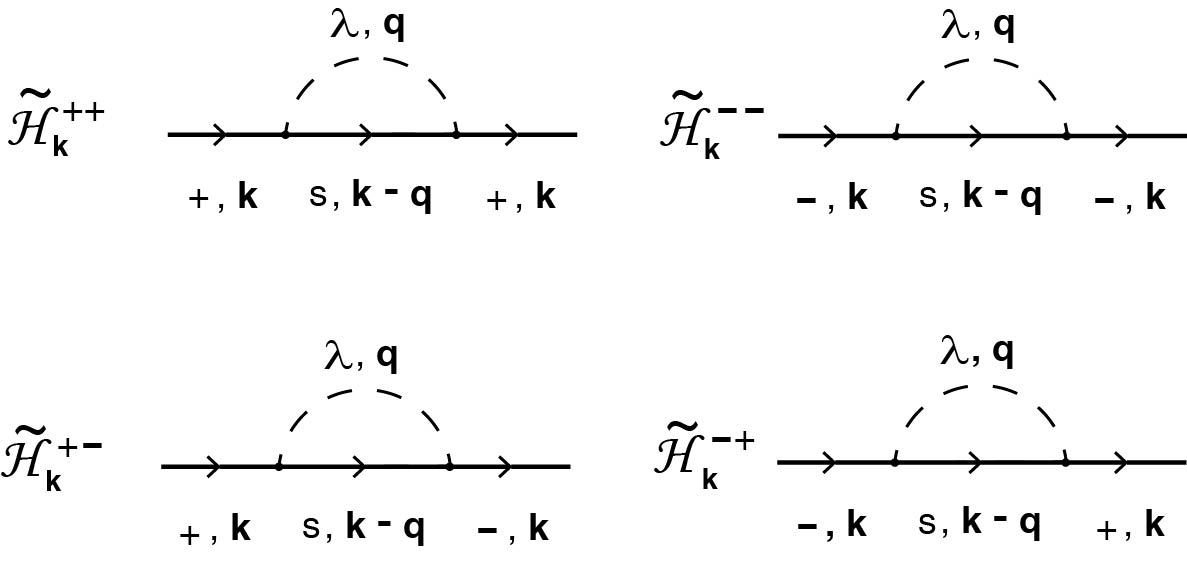}
\caption{Diagrammatic representation of the terms entering in the
Hamiltonian (\ref{eH}). The solid lines correspond to the
electrons and the dashed lines correspond to the virtual photons.
Index $\lambda=\pm$ denotes the two different circular
polarizations of the photons, and the index $s=\pm$ denoted the
two different Dirac cones. Summation should be performed over both
the indices $\lambda$ and $s$.} \label{Fig2}
\end{figure}

Let us briefly remind how the second-order corrections can be
accounted for within the framework of perturbation theory for
degenerate states (the details can be found, e.g., in
Ref.~[\onlinecite{Bir_Pikus}]). Imagine that we have a set of
states $\{m\}$ which are close in energy to each other (this means
that the energy distances
$|\varepsilon_{m}^{(0)}-\varepsilon_{m'}^{(0)}|$ between them are
comparable or smaller relative to a characteristic energy of the
perturbation). The perturbation does not couple any states $m$ and
$m'$ directly (otherwise the standard first-order perturbation
theory is applicable), but couples them to a set of the states
$\{l\}$ whose energies lie far from energies of the states $\{m\}$
(this means that the energy distances
$|\varepsilon_{m}^{(0)}-\varepsilon_{l}^{(0)}|$ are large as
compared with the characteristic energy of the perturbation). In
our case the set $\{m\}$ consists of the two states
$\{|\mathbf{k},+,0,0\rangle,\,|\mathbf{k},-,0,0\rangle\}$, and the
set $\{l\}$ corresponds to the states
$|\mathbf{k}',\pm,N_{+,n,\textbf{q}},N_{-,n,\textbf{q}^\prime}\rangle$
with $N_{+,n,\textbf{q}}+N_{-,n,\textbf{q}^\prime}\neq0$. Then
energies of the perturbed $\{m\}$ states can be obtained by
diagonalization of the matrix Hamiltonian
\begin{equation}
\widetilde{\mathbf{{\cal H}}}=\widetilde{\mathbf{{\cal
H}}}^{(0)}+\widetilde{\mathbf{{\cal H}}}^{(2)}, \label{SHPert}
\end{equation}
where $\widetilde{\mathbf{{\cal H}}}^{(0)}$ is the matrix of
unperturbed Hamiltonian (\ref{HK}) written in the subspace of
states $\{m\}$, and the matrix elements of the Hamiltonian
$\widetilde{\mathbf{{\cal H}}}^{(2)}$ can be found as
\begin{eqnarray}
\widetilde{{\cal H}}^{(2)}_{mm'}&=&\frac{1}{2}\sum_{l}
\left(\frac{1}{\varepsilon_{m}^{0}-\varepsilon_{l}^{0}}+\frac{1}{\varepsilon_{m'}^{0}-\varepsilon_{l}^{0}}\right)\nonumber\\
&\times&\langle m|\hat{{\cal H}}_{\mathrm{int}}|l\rangle\langle
l|\hat{{\cal H}}_{\mathrm{int}}|m'\rangle\,, \label{SHgen}
\end{eqnarray}
where the summation goes over all set of the states $\{l\}$. In
the case we consider, the Hamiltonian $\widetilde{\mathbf{{\cal
H}}}^{(2)}_\textbf{k}$ for a given electron wave vector
$\textbf{k}$ can be written as the $2\times 2$ matrix
\begin{eqnarray}\label{eH}
\widetilde{{\cal H}}^{(2)}_\textbf{k}=\left(\begin{array}{cc}
\hbar v_Fk + \widetilde{{\cal H}}^{++}_\textbf{k} & \widetilde{{\cal H}}^{+-}_\textbf{k} \\
\widetilde{{\cal H}}^{-+}_\textbf{k} & -\hbar v_Fk +
\widetilde{{\cal H}}^{--}_\textbf{k}
\end{array}\right)\,,
\end{eqnarray}
where the vacuum-fluctuation corrections $\widetilde{{\cal
H}}^{++}_\textbf{k}$, $\widetilde{{\cal H}}^{--}_\textbf{k}$ and
$\widetilde{{\cal H}}^{+-}_\textbf{k}=\left(\widetilde{{\cal
H}}^{-+}_\textbf{k}\right)^\ast$ are given by
\begin{widetext}
\begin{eqnarray}\label{Hpp}
\widetilde{{\cal
H}}^{++}_{\textbf{k}}=-\frac{e^2v_F^2}{4\pi\epsilon_{0}
L}\sum_{\lambda=\pm}\int
d^2\textbf{q}\frac{\omega_{\lambda,\textbf{q}}-v_F|\mathbf{k}|}
{\omega_{\lambda,\textbf{q}}\left[(\omega_{\lambda,\textbf{q}}-v_F|\mathbf{k}|)^2-
v_F^2|\textbf{k}-\textbf{q}|^2\right]}\,,
\end{eqnarray}
\begin{eqnarray}\label{Hmm}
\widetilde{{\cal
H}}^{--}_{\textbf{k}}=-\frac{e^2v_F^2}{4\pi\epsilon_{0}
L}\sum_{\lambda=\pm}\int
d^2\textbf{q}\frac{\omega_{\lambda,\textbf{q}}+v_F|\mathbf{k}|}
{\omega_{\lambda,\textbf{q}}\left[(\omega_{\lambda,\textbf{q}}+v_F|\mathbf{k}|)^2-
v_F^2|\textbf{k}-\textbf{q}|^2\right]}\,,
\end{eqnarray}
\begin{eqnarray}\label{Hpm}
\widetilde{{\cal
H}}^{+-}_{\textbf{k}}=-\frac{e^2v_F^2}{4\pi\epsilon_{0}
L}\sum_{\lambda=\pm}\lambda\int
d^2\textbf{q}\frac{\omega_{\lambda,\textbf{q}}^2-v_F^2
(|\mathbf{k}|^2+|\textbf{k}-\textbf{q}|^2)}{\left[(\omega_{\lambda,\textbf{q}}+v_F|\mathbf{k}|)^2-
v_F^2|\textbf{k}-\textbf{q}|^2\right]\cdot\left[(\omega_{\lambda,\textbf{q}}-
v_F|\mathbf{k}|)^2-v_F^2|\textbf{k}-\textbf{q}|^2\right]}\,,
\end{eqnarray}
\end{widetext}
and the symbol $\lambda=\pm$ corresponds to the two different
circular polarizations of virtual photons. The physical meaning of
the terms in the Hamiltonian (\ref{eH}) is the following. The
matrix element $\widetilde{\mathbf{{\cal H}}}^{++}_\textbf{k}$
corresponds to the process: an electron in the upper Dirac cone
emits a virtual photon and then reabsorbs this photon while
returning to the same cone. The matrix element
$\widetilde{\mathbf{{\cal H}}}^{--}_\textbf{k}$ corresponds to the
same process for the electron in the lower Dirac cone. The
off-diagonal matrix elements $\widetilde{\mathbf{{\cal
H}}}^{+-}_\textbf{k}=\widetilde{\mathbf{{\cal
H}}}^{-+}_\textbf{k}$ correspond to the processes in which the
electron after reabsorption of the photon changes the Dirac cone.
Diagrammatic representation of these terms is shown in
Fig.~\ref{Fig2}.
Diagonalization of the Hamiltonian (\ref{eH}) gives the
renormalized energy spectrum of the elementary excitations in
graphene,
\begin{eqnarray}\label{f}
{\varepsilon}^{\pm}(\textbf{k})=\widetilde{{\cal
H}}^{++}_\textbf{k}/2+\widetilde{{\cal
H}}^{--}_\textbf{k}/2\nonumber\\
\pm\sqrt{\left(\widetilde{{\cal
H}}^{++}_\textbf{k}/2-\widetilde{{\cal H}}^{--}_\textbf{k}/2+\hbar
v_F|\mathbf{k}|\right)^2+\left(\widetilde{{\cal
H}}^{+-}_\textbf{k}\right)^2}\,.
\end{eqnarray}
Taking into account that at $\textbf{k}=0$ we have
$\widetilde{{\cal H}}^{++}_\textbf{k} =\widetilde{{\cal
H}}^{--}_\textbf{k}$, the renormalized electron energy (\ref{f})
at the Dirac point can be written as
\begin{equation}\label{f0}
{\varepsilon}^{\pm}(0)=\widetilde{{\cal H}}^{++}_0
\pm\left|\widetilde{{\cal H}}^{+-}_0\right|\,.
\end{equation}
It follows from Eq. (\ref{f0}) that the vacuum fluctuations of
electromagnetic field in the cavity can open the band gap between
the conductivity and valence bands of graphene at the Dirac point,
which is
\begin{equation}\label{G}
\varepsilon_g=2\left|\widetilde{{\cal H}}^{+-}_0\right|\,.
\end{equation}
It should be stressed that in the absence of an optically active
media, the eigenfrequencies of clockwise and counterclockwise
circularly polarized photons are equal,
$\omega_{+,\textbf{q}}=\omega_{-,\textbf{q}}$. According to the
equation (\ref{Hpm}), in this case the term $\widetilde{{\cal
H}}^{+-}_0$ is zero and the band gap (\ref{G}) vanishes. Therefore
for the band gap opening one needs to fill the cavity by an
optically active media which splits modes of virtual photons with
different circular polarizations. In this case the photonic
dispersions read as
\begin{equation}
\omega_{\pm,\textbf{q}}=c_{\pm}\sqrt{q^2+q_{z}^2}
\label{Omegas}\,,
\end{equation}
where $q_z=\pi n/L$ is the quantized $z$-component of photon wave
vector in the cavity, $n$ is the number of photon mode,
$c_\pm=c/n_\pm$ are the speeds of light with clockwise and
counterclockwise circular polarizations, and $n_\pm$ are the
refractive indices for clockwise and counterclockwise polarized
light, which are different in the optically active media, $n_+\neq
n_-$.

In the discussion above we restricted our analysis to the
single-mode approximation, accounting for the coupling of the
electron in graphene with only one photon mode. Going beyond this
approximation, one needs to perform the summation over all modes
$n$ in Eqs. (\ref{Hpp})--(\ref{Hpm}). Keeping in mind that photon
modes with even numbers $n$ correspond to the zero field intensity
in the center of the cavity and, thus, do not interact with the 
graphene sheet, one gets the following expression for the band gap:
\begin{equation}
\varepsilon_g=\frac{e^2}{2\pi\epsilon_{0}L}\left|\sum_{n=0}^{\infty}\sum\limits_{\lambda=\pm}
\frac{\lambda\beta_\lambda^2}{1-
\beta_\lambda^2}\ln\left[1+\frac{q_{0}^2L^2(1-\beta_\lambda^2)}{\pi^2(2n+1)^2}\right]\right|\,,
\label{GapMultimode}
\end{equation}
where $\beta_{\pm}=v_{F}/c_{\pm}$, $q_0\sim 1/a_{0}$ is the
cut-off parameter of integration in Eq. (\ref{Hpm}) with $a_{0}$
being the lattice constant of graphene.

\section{Results and discussion}
It is seen that Eq. (\ref{GapMultimode}) contains the summation
over photon polarizations, $\lambda=\pm$. As expected, the
contributions of clockwise and counterclockwise polarized photons
in the band gap (\ref{GapMultimode}) have opposite signs and the
band gap vanishes for $\beta_+=\beta_-$ and appears only in the
presence of an optically active media with $\beta_+\neq\beta_-$.
For instance, the cavity can be filled with a magneto-gyrotropic
media based on ferrite garnets which possess the giant difference
between the velocities of light with different circular
polarizations.\cite{Ferrite} The effect becomes even more
pronounced if the cavity is filled with an active media with the
circular dichroism.\cite{CD_monography} In this case one of the
two circularly polarized photon modes in the cavity is suppressed
and its contribution to the band gap (\ref{GapMultimode}) can be
neglected. As a result, the summation over $\lambda$ in Eq.
(\ref{GapMultimode}) can be omitted, which leads to the drastic
increasing of the band gap.
\begin{figure}
\includegraphics[width=1.0\linewidth]{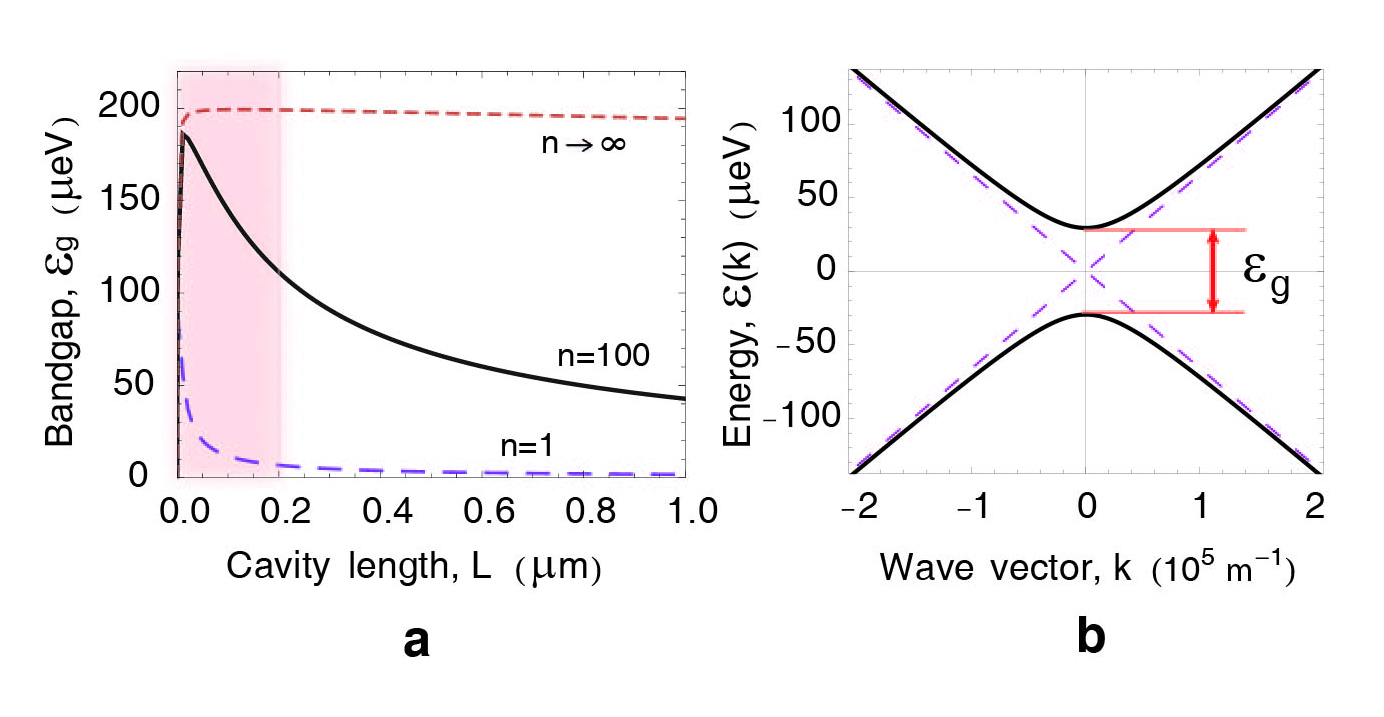}
\caption{(color online) (a) The band gap in graphene induced by
vacuum fluctuations, calculated with accounting different numbers
of cavity modes $n$; (b) Energy spectrum of free electrons in
graphene (dashed lines) and electrons dressed by virtual photons
(solid lines). The calculation is performed for the cavity length
$L=300$ nm and the number of accounted cavity modes is $n=100$.}
\label{Fig3}
\end{figure}

Figure \ref{Fig3}(a) shows the dependence of the band gap,
$\varepsilon_g$, on the cavity length, $L$, for the cavity filled
by such a media with the circular dichroism. In the physically
relevant region of the cavity lengths (the white area in Fig.
\ref{Fig3}(a)), the band gap calculated in the single-mode
approximation ($n=1$) is of several $\mu$eV, which is comparable
with the Lamb shift.\cite{Lamb_47} The summation over higher modes
increases this value by 1--2 orders of magnitude: for 100 modes
the value of the band gap increases to 50--100 $\mu$eV, while the
summation over all modes gives value of about 200 $\mu$eV.
However, the summation over infinite number of modes overestimates
the band gap. Indeed, if the characteristic photon wavelength,
$2\pi/q_z=2L/n$, is comparable to the interatomic distance, the
macroscopic description of an optically active media becomes
irrelevant. Therefore the band gap can be reasonably estimated to
be about tens of $\mu$eV, that is one order of magnitude bigger
than the Lamb shift.\cite{Lamb_47}

The energy spectrum of massive Dirac fermions in graphene is
plotted in Fig. \ref{Fig3}(b). The renormalized dispersion
relation can be approximated by the analytical expression
\begin{equation}\label{s}
{\varepsilon}^{\pm}(\textbf{k})=\pm\sqrt{\left(\hbar v_F
|\mathbf{k}|\right)^2+\left(m^\ast v_{F}^2\right)^2}\,,
\end{equation}
where the effective mass of electron dressed by virtual photons is
$m^\ast=\varepsilon_g/2v_F^2$.

It should be noted that the considered single-electron problem can
be easily generalized for the realistic situation when the valence
band is filled by the Fermi sea of electrons. In this case the
Pauli principle forbids virtual transitions into the lower Dirac
cone filled with electrons, which reduces both the matrix elements
(\ref{Hpp})--(\ref{Hpm}) and the band gap (\ref{GapMultimode}) by
the factor of $1/2$.

\section{Conclusions} We predicted the quantum
electrodynamical effect in graphene placed inside a planar cavity
filled by an optically active media. Due to the vacuum
fluctuations of electromagnetic field in the cavity, the spectrum
of elementary excitations in graphene undergoes qualitative
changes. Namely, the valence and conductivity bands of graphene
are split at the Dirac points. The value of the vacuum-induced
band gap can be one order of magnitude bigger then the famous Lamb
shift in hydrogen atom.

\textit{Acknowledgements.} The work was partially supported by
Rannis ``Center of Excellence in Polaritonics'', Eimskip
foundation, the RFBR projects 10-02-00077 and 10-02-90001, the
Russian Ministry of Education and Science, the 7th European
Framework Programme (Grants No. FP7-230778 and FP7-246784), and ISTC Project No.
B-1708.

\appendix
\section{Derivation of interaction Hamiltonian}
The introduction of the electron-photon interaction in graphene
can be done by the conventional replacement
$\hbar{\textbf{k}}\rightarrow\hbar{\textbf{k}}-e\hat{\textbf{A}}$,
where $\hat{\textbf{A}}$ is the operator of the vector-potential
of electromagnetic field. Then the full Hamiltonian of the
electron-photon system reads
\begin{equation}
\hat{{\cal
H}}=v_F\hat{\bm{\sigma}}\cdot\left(\hbar{\textbf{k}}-e\hat{\textbf{A}}\right)+
\frac{1}{2}\int
dV\left(\epsilon_0\hat{\textbf{E}}^\dagger\hat{\epsilon}\hat{\textbf{E}}+\mu_0\hat{\textbf{B}}^\dagger\hat{\mu}^{-1}\hat{\textbf{B}}\right)
\label{SHamiltonian}\,,
\end{equation}
where $\hat{\textbf{E}},\,\hat{\textbf{B}}$ are the operators of
electric and magnetic fields, and $\hat{\epsilon},\,\hat{\mu}$ are
the tensors of electric and magnetic permittivity of the media,
respectively. The integration in the last term, giving the energy
of free electromagnetic field, goes over all space where the field
is present. In the current paper we consider a graphene sheet
placed in a planar microcavity. In this case it is convenient to
represent the operators of the fields in terms of the eigenmodes
of the cavity as
\begin{eqnarray}
\hat{\mathbf{A}}(\textbf{r})=\sum\limits_{\lambda,n,\textbf{q}}\hat{\mathbf{A}}_{\lambda,n,\textbf{q}}(\textbf{r})\,,\label{S1}\\
\hat{\mathbf{E}}(\textbf{r})=\sum\limits_{\lambda,n,\textbf{q}}\hat{\mathbf{E}}_{\lambda,n,\textbf{q}}(\textbf{r})\,,\label{S2}\\
\hat{\mathbf{B}}(\textbf{r})=\sum\limits_{\lambda,n,\textbf{q}}\hat{\mathbf{B}}_{\lambda,n,\textbf{q}}(\textbf{r})\,,\label{S3}
\end{eqnarray}
where $n=1,2,3,...$ is the number of field mode in the cavity.
Using the Coulomb gauge, we can write the field operators
(\ref{S1})--(\ref{S3}) as
\begin{eqnarray}\nonumber
\hat{\mathbf{A}}_{\lambda,n,\textbf{q}}(\textbf{r})=\sqrt{\frac{\hbar}{2\epsilon_{0}\omega_{\lambda,n,\mathbf{q}}}}
\Big(\hat{a}_{\lambda,n,\mathbf{q}}\mathbf{u}_{\lambda,n,\mathbf{q}}(\textbf{r})+\\
+\hat{a}^{\dagger}_{\lambda,n,\mathbf{q}}\mathbf{u}^{*}_{\lambda,n,\mathbf{q}}(\textbf{r})\Big)
\label{SA},\\ \nonumber
\hat{\mathbf{E}}_{\lambda,n,\textbf{q}}(\textbf{r})=i\sqrt{\frac{\hbar\omega_{\lambda,n,\mathbf{q}}}{2\epsilon_{0}}}
\Big(\hat{a}^{\dagger}_{\lambda,n,\mathbf{q}}\mathbf{u}^{*}_{\lambda,n,\mathbf{q}}(\textbf{r})-\\
-\hat{a}_{\lambda,n,\mathbf{q}}\mathbf{u}_{\lambda,n,\mathbf{q}}(\textbf{r})\Big)
\label{SE},\\ \nonumber
\hat{\mathbf{B}}_{\lambda,n,\textbf{q}}(\textbf{r})=\sqrt{\frac{\hbar}{2\epsilon_{0}\omega_{\lambda,n,\mathbf{q}}}}
\Big(\hat{a}_{\lambda,n,\mathbf{q}}\nabla\times\mathbf{u}_{\lambda,n,\mathbf{q}}(\textbf{r})+\\
+\hat{a}^{\dagger}_{\lambda,n,\mathbf{q}}\nabla\times\mathbf{u}^{*}_{\lambda,n,\mathbf{q}}(\textbf{r})\Big)\,,
\end{eqnarray}
where $\mathbf{u}_{\lambda,n,\mathbf{q}}$ are the cavity
eigenmodes. If the cavity is filled with an optically active
media, the eigenmodes are circularly polarized and can be found as
\begin{eqnarray}\label{Su}
\mathbf{u}_{\pm,n,\mathbf{q}}(z,\mathbf{r})=\mathbf{e_{\pm}}\sqrt{\frac{2}{LS}}\sin\left({\frac{\pi
n z}{L}}\right)e^{i\mathbf{q\cdot r}}\,,
\end{eqnarray}
where
\begin{eqnarray}
\omega_{\pm,n,\mathbf{q}}=c_\pm\sqrt{q^2+\left(\frac{\pi
n}{L}\right)^2} \label{SOmegaPM}
\end{eqnarray}
are the photon eigenfrequencies. Therefore the Hamiltonian of the
interaction between the graphene sheet and the electromagnetic
field in the cavity can be written as
\begin{eqnarray}
\hat{{\cal H}}_{\mathrm{int}} =
-ev_F\hat{{\sigma}}\cdot\hat{\textbf{A}}(\textbf{r})&=&\nonumber
ev_F\sqrt{2}\sum_{\lambda=\pm,n,\mathbf{q}}(\mathbf{e}_{+}\hat{\sigma}^{-}+\\
&+&\mathbf{e}_{-}\hat{\sigma}^{+})
\cdot\hat{\textbf{A}}_{\lambda,n,\mathbf{q}}(\textbf{r}).
\label{SHint}
\end{eqnarray}
Since the graphene sheet is placed in the center of the cavity (at
$z=L/2$), it is coupled only with modes (\ref{Su}) corresponding
to odd numbers $n$. This means that the summation index, $n$, in
Eq.~(\ref{SHint}) is odd: $n=1,3,5,7,\dots$. Then, using the
expression (\ref{SA}) for the vector potential operator
$\hat{\textbf{A}}_{\lambda,n,\mathbf{q}}(\textbf{r})$, the
interaction Hamiltonian (\ref{SHint}) can be rewritten in the form
(\ref{H0}).

\end{document}